# Ultraviolet direct absorption microscopy for single particle protein/nucleic acid quantification

C. J. Richards,[1,*] D. van de Lockand,[1] D. Wolters,[1] and M. Liebel[1,$]

[1]Department of Physics and Astronomy, Vrije Universiteit Amsterdam, De Boelelaan 1100, Amsterdam 1081 HZ, The Netherlands.

[*] c.j.richards@vu.nl
[$] m.liebel@vu.nl

## Abstract

Bio-nanoparticles are pivotal to next generation nanotherapeutics, but providing single-particle biomolecular characterization remains a crucial challenge. Herein we present ultra-violet direct absorption microscopy (UV-DAM) to tackle this challenge. UV-DAM is based on a tailored illumination scheme for absorption-only imaging, combined with a custom deep UV light source. Combined, they provide biomolecular specificity with single particle sensitivity. As such, UV-DAM provides rapid, label-free, high resolution, biochemical imaging. Enabled by these capabilities, we implement the classic nucleic acid:protein absorption assay at the single virus level where we demonstrate UV-DAM's ability to distinguish empty from DNA-loaded viral capsids based on bimolecularly specific absorption fingerprints. UV-DAM presents the translation from bulk UV-visible spectrometry to single-particle assessment, a crucial advancement for nanomedicine characterization where particle loading efficiencies are often heterogenous. Beyond this, UV-DAM is applicable to a wide range of nanomaterials or investigation of biological process with high spatio-temporal resolution and intrinsic molecular contrast.

## Introduction

Bio-nanoparticles are cornerstones of modern medicine - as gene therapy vectors, mRNA carriers, vaccines and diagnostic agents that combine engineered specificity with natural biocompatibility. [1–7] Several are already clinically approved, spanning virus-based[8,9] and lipid-based[6,7,10] formulations. Overall, so-called nanodrugs are poised to define the next generation of therapeutics.[2–5] Yet unlike small-molecule drugs, nanodrugs are poorly-compatible with standard analytical characterization: particle-to-particle variability directly impacts therapeutic efficacy but is notoriously difficult to quantify.[11] Existing tools such as cryo-electron microscopy, atomic force microscopy, flow cytometry, analytical ultracentrifugation, and mass spectrometry are either low-throughput, bulk, or require labels.[11–13] Single-particle scattering methods offer label-free solutions at adequate throughput but demand specialized assays and pristine samples.[12,14]

The most direct solution would be to simply measure the absorption spectrum of each individual nanodrug, thus implementing the single-particle equivalent of the ubiquitous bulk workhorse, the NanoDrop.[15] Because biomolecules absorb light in the deep ultraviolet (DUV) with well-defined, distinct, absorption features[16,17] the resulting spectra would directly report on a particles' molecular composition in a straight-forward, label-free, manner. The clearest example is the 260/280 nm absorbance ratio, known as the nucleic acid:protein ratio that underlies decades of bulk nucleic acid quantification via UV-visible spectrometry.[15,16,18]

Yet translating this measurement to the single-nanoparticle level remains unrealized. DUV absorption imaging could be a solution but, despite a long history and clear advantages combining high spatial resolution with[19] large absorption contrast[20,21] and biomolecular specificity[17,20] DUV hyperspectral imaging for 260/280 nm quantification remains limited to bulk objects such as eukaryotic cells.[22–31] Notably, Fang *et al.* recently applied DUV extinction imaging to metal and plastic nanoparticles, establishing that DUV nanoscopy is conceptually viable but without providing absorption-specificity or the spectral tunability required for genuine biomolecular sensing.[21,32]

In this work we implement a nanoscale "NanoDrop" by introducing DUV direct absorption microscopy (UV-DAM). UV-DAM directly records quantitative absorption signals of many individual nanoparticles at three crucial wavelengths: 260/280/300 nm thus reproducing the established bulk nucleic acid:protein quantification with single-particle sensitivity. UV-DAM is based on a tailored and spectrally tunable DUV light source combined with a custom DUV microscope that favorably enhances DUV absorption contrast through spatial coherence engineering.

The paper is organized as follows. We first discuss the experimental implementation of UV-DAM followed by thorough validation using Au nanoparticles in combination with scanning electron microscopy (SEM) imaging. We then move to specifically chosen biological nanoparticles: bacteriophages, or "phages". We address bioparticle characterization by initially measuring the phage contrast values and demonstrate that photodamage is no concern for low-light UV-DAM. Finally, we implement the iconic nucleic acid:protein quantification scheme on the single virus level by interrogating nucleic acid loaded bacteriophages alongside bacteriophage fractions that we actively unload by chemically triggering nucleic acid ejection.

## Results

<u>Absorption imaging via UV-DAM</u>

Figure 1a schematically outlines the implementation of UV-DAM. A tailored broadly tunable DUV source provides spatially incoherent light from 250–330 nm through a multimode fiber (Materials and Methods). The fiber output is relay imaged into the sample plane of the microscope which is operated in a back-reflection geometry using a 1.2 numerical aperture (NA) glycerol immersion objective (Materials and Methods). Partially reflective Si-substrates ensure that absorption measurements are feasible even for particles located on the surface. A core aspect of UV-DAM is reducing signal-contributions from scattering as biomolecularly specific absorption signatures are of interest. Overfilling the illumination back-focal-plane achieves this goal by essentially reducing the optical transfer function for scattering to zero.[33,34]

To understand this aspect it is instructive to consult object transfer theory and in particular the weak object framework.[35] It describes how the real (absorption) and imaginary (scattering) components of a field are transferred through the imaging pupil for a given source and sample. Figure 1b summarizes the analysis, comparing real and imaginary parts of the transfer function for a typical brightfield imaging configuration with a coherence parameter $s = NA_{ill} / NA_{img} = 0.5$. At perfect focus the absorption signal is maximally transferred, whereas the scattering signal is not transferred for any spatial frequency. However, defocusing leads to transfer of scattering signal, even at small distances of 10 nm. In practice perfect focus is unlikely and samples are three dimensional in nature, thus a component of the scattering contrast will remain in the total measured signal. However, when $NA_{ill} = NA_{img}$ ($s = 1$) the imaginary component of the transfer function is negligible irrespective of defocus, *i.e.* scattering contrast is eliminated (Figure 1b, right). UV-DAM ensures $NA_{ill} = NA_{img}$ by filling the objective pupil which conveniently also maximizes spatial resolution.

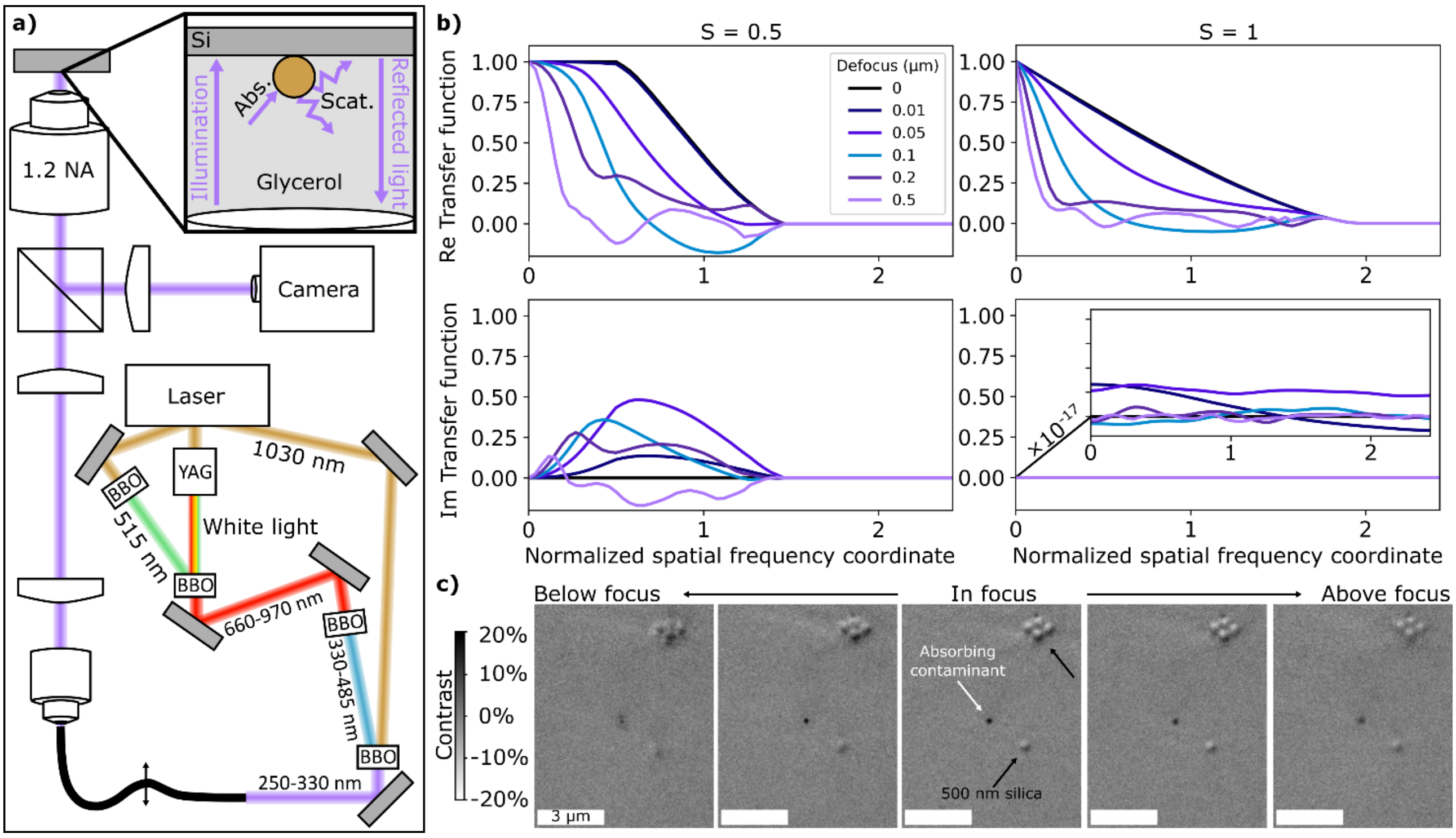

**Figure 1. Microscope and absorption imaging.** a) Experimental implementation of UV-DAM: imaging system and source. Tunable DUV light is generated through several non-linear processes (Materials and Methods). The light is coupled into the microscope via a multimode fiber which is oscillated to reduce speckle. The 4f relay configuration is designed so that the pupil of the objective is fully filled. The microscope operates in a reflection geometry whereby the objective is used to illuminate the sample, and then the field that interacts with the sample is back-reflected from the Si wafer into the same objective. Altogether, this configuration has $NA_{ill} = NA_{img}$, *i.e.* the coherence

parameter $s = 1$. b) Transfer functions for a $s = 0.5$ (left) and $s = 1$ (right) system. Top graphs show the real component of the transfer function (absorption) and the bottom graphs show the imaginary component (scattering) for various defocus distances. The spatial frequency coordinate is normalized to the coherent limit (i.e. a $s = 0$ system). For a $s$ =1 system the imaginary transfer function is on the order of $\times 10^{-17}$ for various defocus distances. c) UV-DAM images of 500 nm diameter $SiO_2$ beads taken with 257.5 nm illumination. The focus position is apparent as 500 nm $SiO_2$ beads (black arrows) and contaminants (white arrow) show little aberration and well-defined edges. When in focus, the $SiO_2$ particles show little contrast whilst contaminants on the surface are clear. Defocus in either direction only reduces the $SiO_2$ particle contrast, indicating absorption-only imaging is achieved.

We first experimentally validated whether UV-DAM indeed achieves absorption-only imaging. Towards this goal, we used 500 nm diameter silica particles as a test case as they have high DUV scattering cross sections, but comparatively small absorption cross sections, $\sigma_{abs}$. This amounts to >|40|% scattering contrast compared to |1.5–10|% absorption contrast, based on Mie theory calculations using various reports of the optical constants for silica.[36,37] Figure 1c shows UV-DAM images of 500 nm silica particles using 257.5 nm illumination. In focus particles (black arrows) show little signal and are difficult to identify as single objects, though clusters are more visible, suggesting that scattering signals are highly suppressed. To further confirm this, we performed defocus imaging. The silica particles exhibit maximum contrast in focus, in agreement with the object transfer calculations for a $NA_{ill} = NA_{img}$ system (Figure 1b, right). The residual, asymmetric, signal is likely a direct result of a spatially non-uniform DUV intensity distribution in the pupil plane which leads to some residual phase transfer in line with Tian *et al*.[38] Therefore, we conclude that UV-DAM strongly favors absorption over scattering contrast, even when defocused.

UV-DAM of Au nanoparticles

To showcase the performance of UV-DAM we characterized Au nanoparticles of nominally 60 nm and 40 nm diameter. Figure 2a,c show scanning electron microscopy (SEM) images of the Au particles on Si substrates, indicating particle sizes of 74 ± 10 nm and 57 ± 9 nm, respectively (particle size histograms shown in Supplementary Figure S1). Figures 2b,d present corresponding UV-DAM images acquired at 280 nm (absorption histograms shown in Supplementary Figure S1). To allow a direct comparison between the two modalities, we converted both the UV-DAM-measured absorption values and the SEM-measured particle diameters to absorption cross section estimates, $\sigma_{abs}$ (Materials and Methods). Figure 2e,f highlight that the converted DUV cross section data mimics the converted SEM data very well for both particle sizes. Overall, UV-DAM readily images individual nanoparticles with measured absorption values matching theoretical expectations.

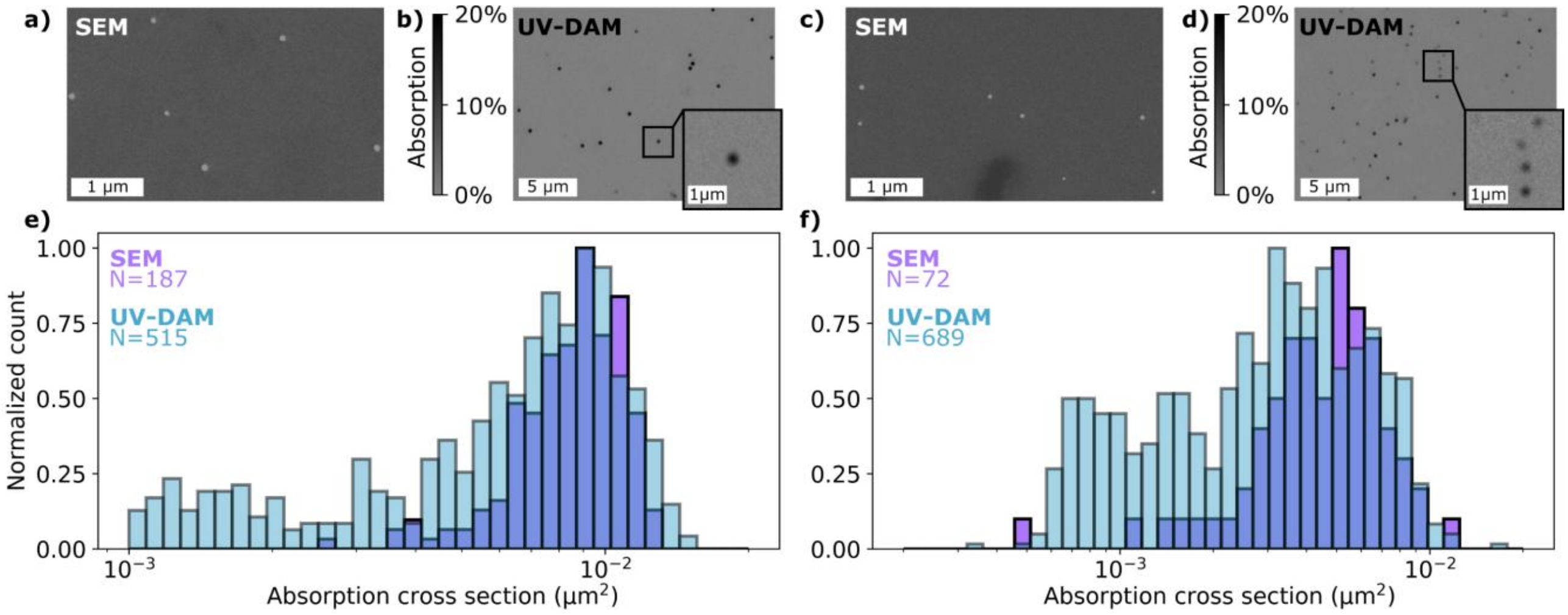

**Figure 2. Characterization of Au nanoparticles.** a) SEM image of nominally 60 nm diameter Au nanoparticles deposited on APTES-functionalized Si wafers and b) UV-DAM image of the same sample. c) SEM measurements of nominally 40 nm Au nanoparticles and d) UV-DAM measurements of the same sample. e, f) Absorption cross section histograms of nominally 60 nm and 40 nm Au nanoparticles, respectively. SEM particle diameters were converted to absorption cross sections using Mie theory and UV-DAM absorption values were converted using Equation 1. The bulk of the SEM (purple) and UV-DAM (blue) distributions are in excellent agreement. There are some discrepancies at low cross section values which are likely to be contaminants that were deposited in the time between SEM and UV-DAM imaging. Note the x-axis log scale. N is the number of particles measured to form the distributions.

## Single virus UV-DAM

Having demonstrated that UV-DAM reproduces SEM-based absorption cross section estimates, we moved towards biologically relevant nanomaterials in the form of bacteriophages. The T5 phage is a siphovirus comprised of an icosahedral protein capsid of approximately 90 nm in diameter encapsulating 120 kbp of condensed viral DNA, and a flexible protein tail of about 250 nm in length (Figure 3a).[39] Due to the hierarchal assembly process, these phages have a highly uniform size distribution, both in terms of capsid diameter as well as nucleic acid content, making them ideal biological reference materials.

Given the prominent standing of holographic imaging approaches, such as iSCAT, in the realm of nanosizing it is instructive to directly compare UV-DAM and iSCAT images of the T5 phages. Figure 3b,c shows such a comparison indicating similarly narrow signal distributions for both modalities. For iSCAT, we observe an average scattering contrast of -3.5 ± 0.3% (Figure 3b) whereas UV-DAM yields an absorption value of 3.4 ± 0.4% (Figure 3c). Both modalities readily identify the particles, with UV-DAM showing a markedly improved spatial resolution due to the shorter illumination wavelength (Figure 3b,c and Supplementary Figure S2). Importantly, the similarity between scattering contrast and absorption value is coincidental. UV-DAM is essentially a brightfield microscope that detects the full illumination light. iSCAT, on the other hand, derives contrast with respect to a dramatically >100-fold attenuated back-reflection. Figure 3d highlights this difference by normalizing the signals with respect to the illumination fluence which we estimate to be 0.1-0.01 W/cm$^2$ for UV-DAM and ~10 W/cm$^2$ for iSCAT. In other words, UV-DAM has a substantial ~150 fold higher normalized signal compared to iSCAT.

These low irradiation levels have important implications as photodamage is often quoted as a major concern in the DUV spectral range[40,41] even though recent work suggests that historical damage estimates might be dramatically overstated.[42] At UV-DAM's low fluences we don't expect photodamage, which we verified via photobleaching experiments on individual phages over extended observation times exceeding 10 min (Figure 3e). We observe a minor signal decrease over minutes, which could also stem from drift in the axial position of the sample rather than photobleaching. Overall, the effect of photobleaching is negligible thus validating the broad applicability of UV-DAM for biological and biomedical nanosizing applications.

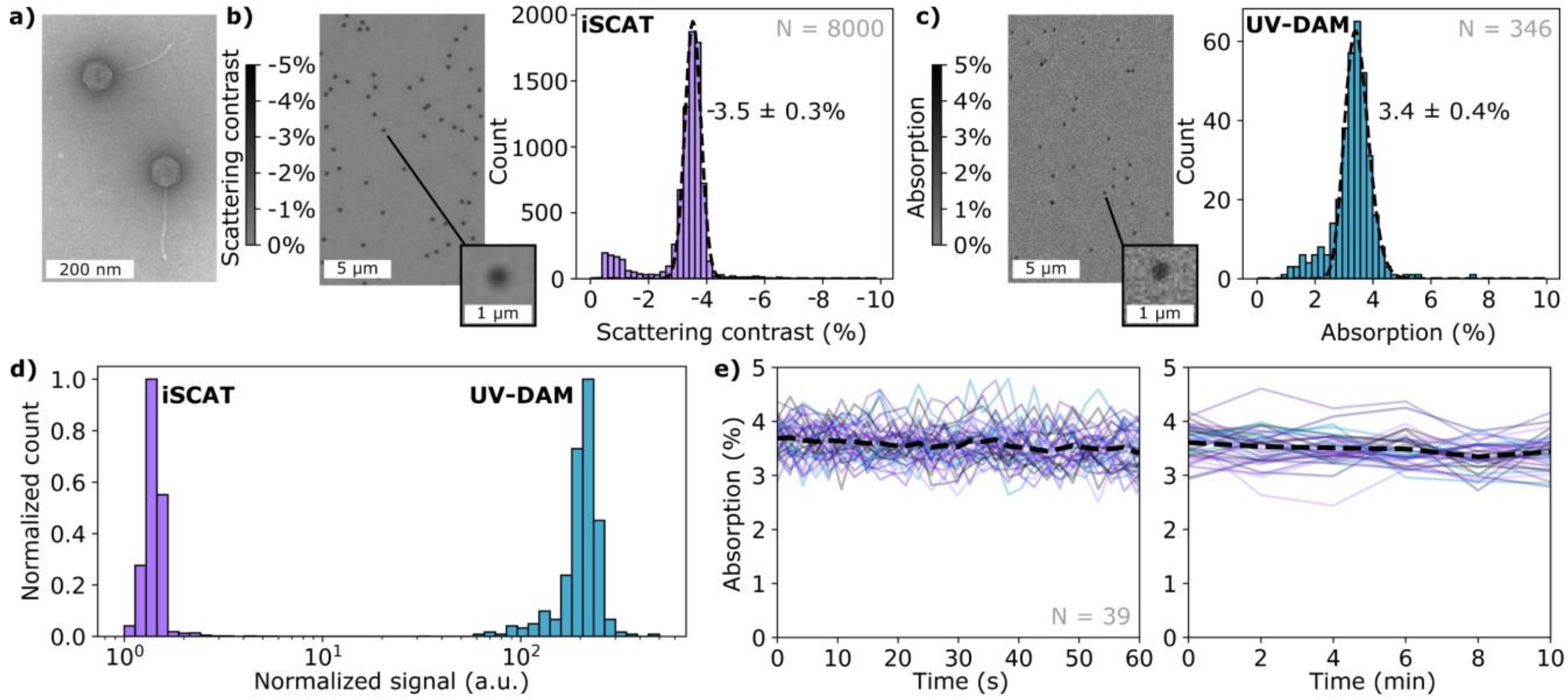

**Figure 3. Imaging bacteriophage particles.** a) Transmission electron microscopy (TEM) image of full T5 bacteriophages with a clear icosahedral capsid and long tail b) iSCAT and c) UV-DAM images of T5 phages shown (left) and corresponding histograms (right). Gaussian or log-normal fits (for iSCAT and UV-DAM data, respectively) of the particle distributions are given by the dashed black lines and corresponding averages and standard deviations are listed. For both techniques phages show single peak distributions. d) Comparison of iSCAT and UV-DAM normalized signals given by contrast × $R$, where $R$ is the wavelength-dependent substrate reflectivity. Note the log scale and log binning. e) Phage photobleaching experiments using 280 nm light performed on short timescales (left) and longer time scales (right). Each solid line is the time dependent absorption of an individual phage, whereas the dashed black line is the average absorption across phages. Negligible photobleaching is observed. N is the number of particles measured.

<u>Biomolecular characterization of phage capsid content</u>

Thus far we have demonstrated that UV-DAM is capable of absorption-only imaging, obtains expected absorption values for metal nanoparticles, and is well-suited for interrogating bio-nanoparticles without photodamage. We now turn to UV-DAM based biomolecular characterization focusing on three specific biomolecular absorption peaks within the DUV spectral region; 260 nm for nucleic acids, 280 nm for proteins, and 300 nm as a general low absorption wavelength.

Figure 4a shows UV-DAM images and absorption histograms of T5 phages using 260 nm, 280nm, and 300 nm illumination (tunable source spectra shown in Supplementary Figure S3). The particles are visible at all illumination wavelengths (further images shown in Supplementary Figure S4) and the histograms show clear single populations. The average absorption values decrease with increasing wavelength, as expected given that ~70% of the total phage mass is DNA (peak absorption at 260 nm) and the remaining ~30% is protein (peak absorption at 280 nm).

Based on the UV-DAM images we can calculate the nucleic acid:protein, or 260/280 nm, absorbance ratio and compare this to values obtained with bulk UV-visible spectrometry. A direct division of the population averages shown in Figure 4a yields a ratio of 1.3, in good agreement with the value of 1.4 obtained from UV-visible spectrometry. Single particle measurements confirm the bulk-results with individual particle absorbance ratios exhibiting values of 1.4 ± 0.4 (Supplementary Figure S5).

As a final step towards biomedical applications, we focused on distinguishing phages with different DNA content. T5 phages eject the 120 kbp of encapsulated viral DNA when incubated with their receptor, FhuA, as schematically shown in Figure 4b.[43,44] The empty phages have a substantially reduced mass and a different biomolecular identity compared to the full phages; DNA + protein for full phages and only protein for empty phages. iSCAT measurements indicated that the phages incubated with FhuA successfully resulted in a mixed sample comprised of both full and empty phages (Supplementary Figure S6).

Figure 4c shows UV-DAM images obtained for the mixed T5 samples at 260 nm, 280 nm and 300 nm. Compared to the loaded phage sample (Figure 4a), the 260 nm UV-DAM images of the mixed sample show increased heterogeneity, with apparent lower and higher contrast particles (Figure 4c arrows). The accompanying absorption histogram shows two resolvable populations of particles. The population with higher absorption values are the remaining full phages within the sample which have a similar value (4.5 ± 0.8%) to the full T5 population shown in Figure 4a (4.3 ± 0.7%). The peak at lower absorption values (1.9 ± 0.6%) are empty phages, which absorb substantially less 260 nm light due to the absence of DNA. The remaining contrast is provided by the low but non-zero absorption of the protein capsid and tail.

At 280 nm illumination the absorption histogram shows a similar, yet less pronounced, trend. This observation is in line with the fact that the protein content of full and empty phages is identical and only low level DNA absorption contributes to the differences. As such, we can distinguish the sample heterogeneity even with 280 nm illumination.

Lastly, using 300 nm illumination we obtain a single-peak distribution with an average absorption of 1.8 ± 0.5%, similar to the full phages (Figure 4a). We expect that at this wavelength the empty phages no longer provide enough contrast to be distinguishable from the background. Mie theory calculations show that the absorption cross section at 300 nm of a 90 nm DNA sphere[45] is >15-fold higher than for a 90 nm protein sphere.[46] We anticipate that the protein signal is even lower as the geometry does not reflect the core-shell structure of the phage. Therefore, though 300 nm illumination is a lowered absorption regime for both DNA and proteins, the DNA-containing full phages still provide sufficient contrast. Indeed, UV-DAM images at 260 nm had an average density of 0.46 ± 0.1 detected objects/$\mu m^2$ and 0.29 ± 0.05 /$\mu m^2$ at 300 nm, *i.e.* ~37% of objects disappeared at 300 nm, corresponding well to the ~42% empty phage population observed at 260 nm and in line with our notion of absorption-only imaging. Overall, we achieve biomolecular sensitivity by using various DUV wavelengths and can distinguish variations in capsid composition.

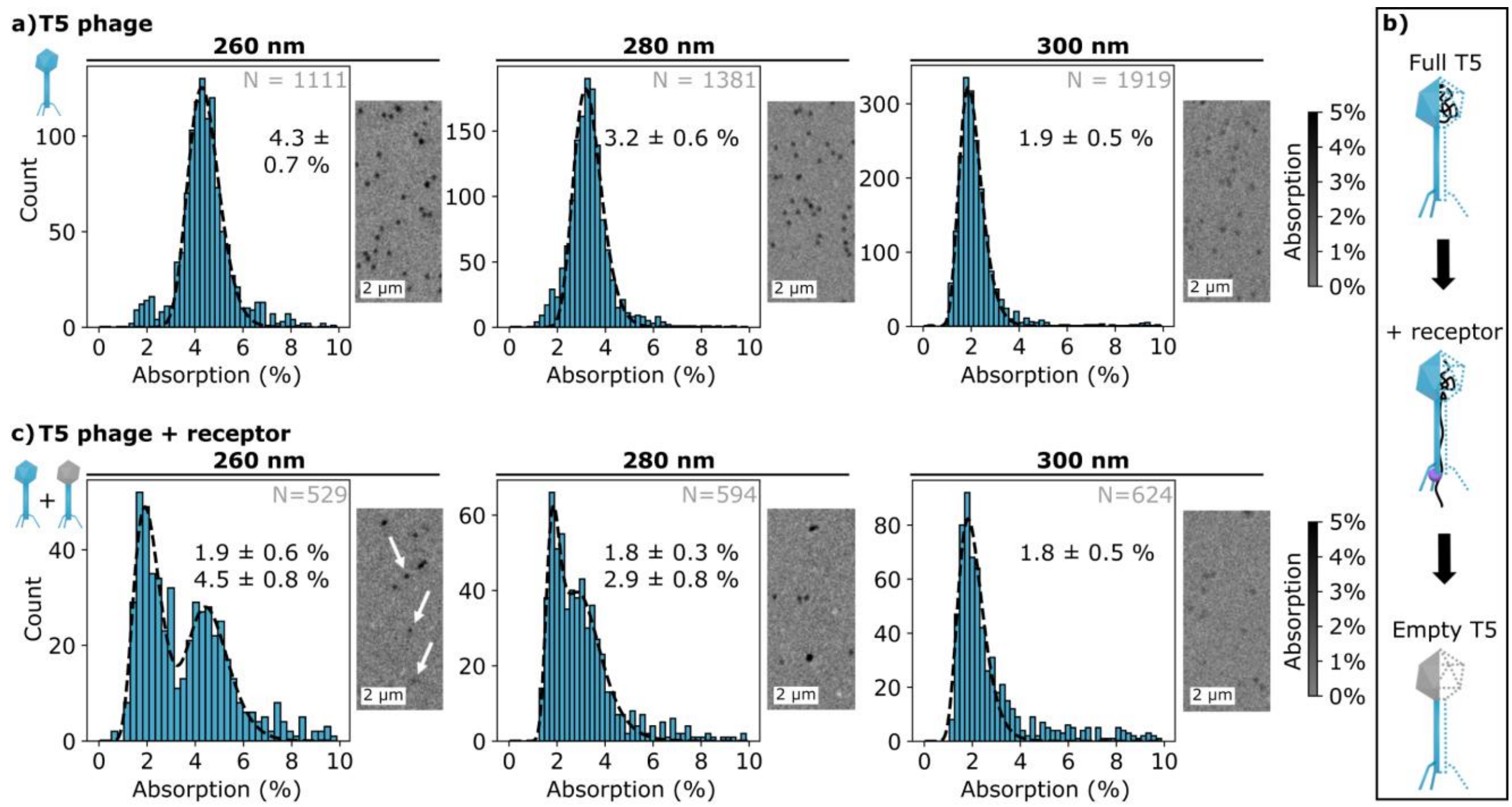

**Figure 4. Hyperspectral biomolecular characterization of phages.** a) DUV hyperspectral measurements of T5 phages performed with 260 nm, 280 nm and 300 nm illumination. Images of the phage samples are presented next to the histograms. The absorption histograms show a single narrow distribution at all wavelengths in accordance with a single particle type; full capsids. b) Schematic representation of T5 bacteriophages. Full phages (blue capsid) eject DNA upon receptor binding, resulting in empty (grey) capsids. c) DUV hyperspectral measurements of T5 phages incubated with receptor. Two populations of objects are observed when using 260 nm illumination (arrows) and to a lesser degree at 280 nm illumination. The histograms show two peaks; empty and full phages with lower and higher absorption, respectively. At 300 nm the samples with and without receptor appear similar. Fits of the particle absorption distributions are given by the dashed black lines and corresponding averages and standard deviations are listed. N is the number of particles measured to form the distributions.

## Discussion

UV-DAM delivers label-free imaging and sizing capabilities with a focus on molecularly informative absorption characteristics rather than scattering. At its core, UV-DAM is a simple microscope operated with matching illumination and collection numerical apertures to favorably boost absorption over scattering contrast, as validated through both object transfer theory and experiments. For biological applications it is precisely the absorption signal, rather than the scattering signal, that holds relevance in terms of molecular characterization. Moreover, UV-DAM explicitly capitalizes on high numerical aperture illumination to boost the spatial resolution, a considerable advantage over interferometric detection of scattering where high numerical aperture illumination is detrimental to contrast.[14,33]

Though UV-DAM suppresses scattering contributions in favor of spectroscopic insight, the measured signal still surpasses previous implementations that measured full extinction; the sum of absorption and scattering obtained using mismatching illumination and collection apertures.[21] UV-DAM almost doubles the contrast values thus making UV-DAM not only well-suited for easy-to-detect metal objects but also for highly relevant yet weaker absorbing biological nanomaterials. The spectral characterization of the T5 phages underscores these capabilities and further demonstrates the broad applicability of UV-DAM across a wide spectrum of nanomaterials, relevant to various applications.

The performance of UV-DAM invites a direct comparison with iSCAT, the label-free optical standard for high-sensitivity nanoscopy down to the single protein level[47,48] which is increasingly being applied to routine characterization of larger bioparticles such as viruses.[49–52] For iSCAT and UV-DAM alike the measured contrast scales inversely with the square-root of the reflectivity, $R$. $R$ is ~62% for a Si-glycerol interface in the DUV, whereas it is 0.4% for iSCAT's glass-water interface in the visible. The reduced $R$ for iSCAT translates into large contrasts, even for weekly scattering objects, and is key to its exquisite sensitivity.[53] Despite this dramatic difference, UV-DAM recovers large absorption signals comparable to the low-reflectivity boosted iSCAT contrasts. This observation is highly relevant as it underlines the photon-efficiency of UV-DAM which requires an approximately 150-fold lower illumination intensity compared to iSCAT to obtain the same signal magnitudes. In its current implementation, UV-DAM operates at below sunlight fluences enabled by the gigantic absorption cross sections of most biomolecules in the DUV spectral range.

UV-DAM's intrinsically low-light imaging approach opens avenues for biological imaging.[40,41] The fluences used here cause negligible photodamage and, even though live-cell work will require further phototoxicity assessments, our results and recent studies[22,42] suggest that live-cell DUV hyperspectral imaging might be viable. Unlike many super-resolution methods, UV-DAM is fast and label-free, sidestepping the toxicity and structural perturbation that labels can introduce thus positioning it to follow biological processes as they happen, at potentially below 100 nm spatial resolution with intrinsic molecular specificity.

Taken together, UV-DAM transitions the widely used protein:nucleic acid absorption measurement via a bulk spectrometer, to the single particle level on the nanoscale.[15] Our T5 260/280 nm ratio-measurements reproduce population averages on the single-particle level and distinguish full and empty phages based on differences in their nucleic acid content. The latter aspect is a true breakthrough for the characterization of biological and biomedical materials where sample-heterogeneity renders established, bulk, methods ill-suited. The T5 bacteriophage is a simple system that highlights UV-DAM's advantages: the phages either do or do not contain DNA. However in many applications the amount of nucleic acids may differ per particle.[54,55] In bulk approaches such as UV-visible spectrometry, these samples may give similar absorption spectra, despite differences in the population distribution. By using UV-DAM to assess the wavelength-dependent absorbance signals on a single particle level it is possible to distinguish samples that have an all-or-none nucleic acid content, such as the T5 phage, from samples with a range of DNA contents. Beyond the current implementation, the DUV source presented here is spectrally tunable, which extends the range of biomolecular species that can be imaged to include lipids which have an absorption peak at 320 nm.

Taken together, UV-DAM is a straight-forward to implement technology for hyperspectral absorption measurements in the biologically relevant DUV wavelength range. In simple terms, UV-DAM is a single-particle NanoDrop, one of the key-workhorses in any microbiology lab, that is further enhanced with high-resolution imaging capabilities. These features open up several exciting avenues to both explore fundamental biological questions with high spatial resolution and temporal dynamics, and to provide high-quality nanocharacterization for biomedicine and biotechnology. Examples are the infection kinetics of bacteria by bacteriophages or label-free imaging cargo transport and nutrient flows in organisms. Moreover, UV-DAM has applied relevance, such as label-free assessment of the loading efficiency of gene delivery nanodrugs. Given that the same biological building blocks are used by all organisms, UV-DAM is readily extendable beyond bacteriophages, to other virus-based delivery vehicles, extracellular vesicles, or lipid nanoparticles. Therefore, we ultimately envision the development of UV-DAM towards a NanoDrop 2.0 in the form of a standard bio-lab technique capable of fast, high-resolution, molecularly specific, single particle characterization for any nano bio-material of interest.

## Materials and methods

UV-DAM setup and image acquisition

Figure 1a presents a schematic of the UV-DAM implementation. Light from the custom DUV light source was coupled into a 400 μm core fiber (0.22 NA, solarization resistant, Thorlabs) which was oscillated to reduce speckle. The fiber output was then relay imaged into the microscope's image plane by means of an Ultrafluar condenser (Zeiss, 0.8 NA) and a 2-lens relay comprising fused silica lenses (Thorlabs). An Ultrafluar objective (Zeiss, 1.2 NA, glycerol immersion, 248 nm design wavelength) both illuminated and imaged the sample. A thick fused silica window placed between the second relay lens and the microscope objective acted as a beamsplitter for illumination and imaging. An f=100 mm lens formed an image on a HB-8000-SB-U camera (Emergent Vision Technologies). The resulting image magnification was 61× resulting in 45.21 nm/pixel. A 27 × 27 μm field of view was used.

Measurements typically used an exposure time of 40–99 ms and acquisition was performed at 10 Hz. Median background images were taken by median averaging 100 frames whilst moving the sample *x*–*y* position. Data was obtained by integrating 20 frames whilst the sample was stationary. For defocus imaging, the sample stage was positioned in the axial direction using a piezo and controller (piezosystem jena GmbH).

DUV source

The Ytterbium:KGW based tunable DUV source, delivering 250–330 nm light is schematically shown in Figure 1a. In brief, the 1030 nm laser output (Pharos PH2-20W-SP, 1 MHz, 20 μJ, ~150 fs) was split into <1 μJ for whitelight generation with the remainder being frequency doubled in a β-barium oxide (BBO) crystal to obtain ~9 μJ at 515 nm and 10 μJ at 1030 nm. A whitelight seed was generated by focusing the <1 μJ of 1030 nm light into a 10 mm YAG crystal (Eksma) and then temporally chirped using a 15 cm BK7 rod (Edmund Optics). A double-pass, 515 nm pumped, non-collinear optical parametric amplifier then amplified the seed, yielding tunable light in the 660-970 nm range with intensities around 1 μJ. The resulting output was then frequency doubled in a BBO crystal to reach the UV range (330-485 nm). Tunable DUV light was obtained via sum frequency generation between the tunable UV and the residual fundamental using a BBO. A slightly nonlinear configuration was employed to background-free isolate the DUV.

A DUV-capable spectrometer (Avantes) was used to measure the DUV source spectra. The tunable source was not used for imaging the 500 nm $SiO_2$ samples, but instead the 515 nm second harmonic light was frequency doubled again to produce 257.5 nm light.

Sample preparation

Diced Si wafer pieces were cleaned with acetone followed by 5 min oxygen plasma cleaning and subsequent functionalization with (3-aminopropyl)triethoxysilane (APTES, Carl Roth GmbH). In brief, a 10% APTES solution in ethanol was placed in a chamber with Si wafers and incubated for 90 min at 150 °C for gas phase deposition. The wafers were then submerged in a 7% acetic acid solution (Carl Roth GmbH) for 10 min, followed by 3 min in ethanol, and finally rinsed with milli-q water and dried with nitrogen gas. APTES functionalized Si wafers were stored under vacuum until use.

For the Au particle samples, stock solutions in citric acid with nominal diameters of 60 nm and 40 nm (CytoDiagnostics) were diluted 20–50× in milli-q, briefly sonicated, and then deposited on functionalized Si wafers for 2 min. Samples were then washed with milli-q and dried with nitrogen gas. 500 nm $SiO_2$ particle samples were made by 500× dilution of the particle stock (microparticles GmbH) in ethanol. A droplet was placed on unfunctionalized Si wafers and left to evaporate.

For the bacteriophage experiments, both T5 phages and FhuA protein stocks were obtained from Amélie Leforestier, Université Paris-Saclay. To prepare full phage samples, the T5 phage stock solution was diluted to a concentration of 40 pM in phage buffer: 100 mM NaCl, 1mM $MgCl_2$, 1mM $CaCl_2$, 10 mM Tris (Thermo Fisher Scientific), and 0.03% N,N-Dimethyldodecylamine N-oxide (MedChemExpress). A droplet of diluted phage suspension was placed on APTES-functionalized Si wafers for 2 min and then rinsed with phage buffer and dried with nitrogen gas. The ejected phage samples were produced by adding FhuA protein to T5 phages at a 1000:1 FhuA:T5 ratio. DNase I (Thermo Fisher Scientific) was added for a final concentration of 0.014 U/µl to digest any ejected DNA, and the sample was incubated at 40 °C for 2 h. The ejected sample was diluted 2 fold in buffer for a final phage concentration of 40 pM and deposited on Si wafers as described above. Samples for iSCAT measurements were similarly prepared and then the phages in solution were deposited on APTES functionalized glass coverslips (No.1.5#, Thorlabs). The solution was left on the sample throughout iSCAT measurement.

Absorption image analysis

UV-DAM images were divided by corresponding median background images and then normalized to the mean being around 1 to account for temporal fluctuations in the beam intensity. Particle detection was performed in python (V3.12). In brief, difference of Gaussian filtering was used to find local maxima. Only one local maxima was permitted per neighborhood size of 10 pixels. Subareas around these co-ordinates in the unfiltered images were then fit with two dimensional Gaussians. The particle absorption values were given by the amplitude of the Gaussian fit. We note that absorption only results in loss of signal, and thus all values are quoted as absolute values. Distributions of the particle absorption values were fit with a single log-normal distribution, or double log-normal distribution when two peaks were present. The stated values are the median of the log-normal distribution, $e^{\mu}$, and the converted standard deviation given by Equation (1) where $\mu$ and $\sigma$ are the mean and standard deviation of the distribution in log-space.

$$STD = \sqrt{e^{\sigma^2-1} \times e^{2\mu+\sigma^2}} \quad (1)$$

Theoretical values for the absorption and scattering cross sections of Au and $SiO_2$ particles were calculated using Mie theory with open source code and using the nominal diameters for $SiO_2$ particles and the SEM measured diameters for the Au particles.[56,57] Refractive index values of $SiO_2$[36,37], Au,[58] glycerol[59], DNA,[45] and the amino acid tyrosine[46] were obtained from the quoted references. The cross sections, $\sigma_{abs}$, were converted into expected contrasts using Equation (2) where the resolution was given by $\lambda_{DUV}$ / (2 × NA). Likewise, UV-DAM measured contrasts were converted to absorption cross sections by inverting Equation (2).

$$Contrast = \frac{\sigma_{abs}}{\pi \times resolution^2} \quad (2)$$

260:280 nm absorbance ratios for single particles were obtained by first spatially aligning hyperspectral images of a region of interest. Particles in the 280 nm illumination image were identified, and then 10 pixel boxes around the identified particle co-ordinates were used to identify the same particles in the 260 nm illumination image. Negative ratio values and those above a value of 3 were removed from the data set as these likely came from issues in particle identification, fitting, or contaminants.

<u>Object transfer theory</u>

Object transfer calculations were performed using the weak object approximation as described in detail in reference.[35] This was compared to calculations using the Abbe method,[38] and results were the same aside from some minor rounding errors due to limited machine number precision. In brief, the weak object transfer function, *WOTF*, is given by Equation 3:

$$WOTF(\boldsymbol{u}) = \int\int S(\boldsymbol{u}')P(\boldsymbol{u}' + \boldsymbol{u})P^*(\boldsymbol{u}')d\boldsymbol{u}' \quad (3)$$

where, *S* is the intensity distribution of the light source, *P* is the pupil function, and $\boldsymbol{u}$ is the two-dimensional coordinate in Fourier space corresponding to the real space coordinate $\boldsymbol{x}$. The real and imaginary (absorption and scattering, respectively) parts of the transfer function are given by:

$$H_A(\boldsymbol{u}) = 2a_0 Re[WOTF(\boldsymbol{u})] \quad (4)$$

$$H_S(\boldsymbol{u}) = -2a_0^2 Im[WOTF(\boldsymbol{u})] \quad (5)$$

where $a_0$ is the mean of the absorption distribution. The calculation was performed using an illumination wavelength of 257.5 nm, the refractive index of glycerol at 257.5 nm,[59] an image pixel size of 45.21 nm (equal to the experiments), and an image size of 128 pixels. The amplitude of the pupil function was made with a lens function with NA= 1.2 (same as the objective used in the setup). Propagation in the axial direction was performed using the angular spectrum method, introducing a phase component to the pupil function. The amplitude distribution of the source-pupil was a lens function with NA equal to *s* × 1.2 where the coherence parameter, *s*, was either 0.5 or 1. The *WOTF*, $H_A(\boldsymbol{u})$, and $H_S(\boldsymbol{u})$ were calculated according to Equations 3–5.

<u>iSCAT</u>

A home-built iSCAT setup based on 470 nm LED light coupled into a fiber (NA=0.22, 550 µm core, Thor labs) was used. The fiber was relay imaged into the sample plane via a 10×/0.25 NA objective (Olympus) and a set of two plano convex lenses followed by a polarizing beam splitter and quarter waveplate and the imaging objective (Olympus, 60×, 1.49 NA, oil immersion). A 300 mm achromatic doublet then formed the polarization rotated iSCAT image on a A BFS-U3-17S7M-C camera (Teledyne Vision Solutions, pixel size of 9 µm). The resultant magnification was 100×, yielding an image pixel size of 90 nm. An integration time of 8 ms per frame was used with an imaging rate of 180 Hz. 6000 frames were median averaged to form median background images, whereas the data was obtained by summing 50 frames. Median background images were divided from the data to give relative scattering contrast values. Images were increased in size from 600 to 1024 pixels by Fourier-domain zero padding interpolation. Particles were detected using difference of Gaussian filtering. The particles were then fit with a two dimensional Gaussian to yield the scattering contrast values which may take positive or negative values, where negative values represent fewer measured photons compared to the background. Distributions of the particle scattering contrasts were fit with a Gaussian and the stated values are the corresponding means and standard deviations of the fit.

Electron microscopy

SEM images of the same samples used for UV-DAM measurements were taken with a FEI Verios 460. Particles size distributions were fit with a gaussian and the stated values are the mean and standard deviation. For TEM, phage preparations were spotted on freshly glow-discharged carbon/formvar-coated mesh grids. After blotting off the excess liquid, the samples were contrasted by 2% uranylacetate (Polysciences Inc, Cat No 21447-25) in water for 1 minute, the excess stain was blotted off and grids were air-dried. Phages were imaged on a 100 kV Tecnai 12 (ThermoFisher) TEM using a 2k x 2k pixel CCD side-mounted camera (Veleta, EMsis GmbH).

## Acknowledgements

The authors would like to acknowledge Amélie Leforestier and Laetitia Gargowitsch (Laboratoire de Physique des Solides, Université Paris-Saclay) for providing the T5 phage and FhuA protein, as well as advice on ejection experiments. We would likewise like to thank Jan Weering and Rien Dekker of the Department of Functional Genomics at the Vrije Universiteit Amsterdam for the TEM imaging, AMOLF Nanolab for access to the SEM, and Sven Askes of the PhotoConversion Materials department at the Vrije Universiteit Amsterdam for access to the UV-visible spectrometer.

## Author contributions

ML conceptualized and supervised the project, and built the DUV light source. ML and CJR constructed the UV-DAM system. CJR performed experiments and analysis. DW programmed the camera. DvdL and ML performed DUV source tuning for hyperspectral measurements. CJR and ML wrote the manuscript.

## Declaration of interests

The authors have no interests to declare

## Supplementary

The supplementary materials include: Au particle histograms, phage line profiles, DUV source spectra, additional hyperspectral images of T5 phages, single particle absorbance ratios, and iSCAT measurements of ejected phage samples.

## References

(1) Shan, X.; Gong, X.; Li, J.; Wen, J.; Li, Y.; Zhang, Z. Current Approaches of Nanomedicines in the Market and Various Stage of Clinical Translation. *Acta Pharmaceutica Sinica B* **2022**, *12* (7), 3028–3048. DOI:10.1016/j.apsb.2022.02.025.

(2) Xu, L.; Wang, X.; Liu, Y.; Yang, G.; Falconer, R. J.; Zhao, C.-X. Lipid Nanoparticles for Drug Delivery. *Advanced NanoBiomed Research* **2022**, *2* (2), 2100109. DOI:10.1002/anbr.202100109.

(3) Herrmann, I. K.; Wood, M. J. A.; Fuhrmann, G. Extracellular Vesicles as a Next-Generation Drug Delivery Platform. *Nat. Nanotechnol.* **2021**, *16* (7), 748–759. DOI:10.1038/s41565-021-00931-2.

(4) Gordillo Altamirano, F. L.; Barr, J. J. Phage Therapy in the Postantibiotic Era. *Clinical Microbiology Reviews* **2019**, *32* (2), 10.1128/cmr.00066-18. DOI:10.1128/cmr.00066-18.

(5) Huang, X.; Kong, N.; Zhang, X.; Cao, Y.; Langer, R.; Tao, W. The Landscape of mRNA Nanomedicine. *Nat Med* **2022**, *28* (11), 2273–2287. DOI:10.1038/s41591-022-02061-1.

(6) Akinc, A.; Maier, M. A.; Manoharan, M.; Fitzgerald, K.; Jayaraman, M.; Barros, S.; Ansell, S.; Du, X.; Hope, M. J.; Madden, T. D.; Mui, B. L.; Semple, S. C.; Tam, Y. K.; Ciufolini, M.; Witzigmann, D.; Kulkarni, J. A.; van der Meel, R.; Cullis, P. R. The Onpattro Story and the Clinical Translation of Nanomedicines Containing Nucleic Acid-Based Drugs. *Nat. Nanotechnol.* **2019**, *14* (12), 1084–1087. DOI:10.1038/s41565-019-0591-y.

(7) Chung, Y. H.; Beiss, V.; Fiering, S. N.; Steinmetz, N. F. COVID-19 Vaccine Frontrunners and Their Nanotechnology Design. *ACS Nano* **2020**, *14* (10), 12522–12537. DOI:10.1021/acsnano.0c07197.

(8) Day, J. W.; Finkel, R. S.; Chiriboga, C. A.; Connolly, A. M.; Crawford, T. O.; Darras, B. T.; Iannaccone, S. T.; Kuntz, N. L.; Peña, L. D. M.; Shieh, P. B.; Smith, E. C.; Kwon, J. M.; Zaidman, C. M.; Schultz, M.; Feltner, D. E.; Tauscher-Wisniewski, S.; Ouyang, H.; Chand, D. H.; Sproule, D. M.; Macek, T. A.; Mendell, J. R. Onasemnogene Abeparvovec Gene Therapy for Symptomatic Infantile-Onset Spinal Muscular Atrophy in Patients with Two Copies of SMN2 (STR1VE): An Open-Label, Single-Arm, Multicentre, Phase 3 Trial. *The Lancet Neurology* **2021**, *20* (4), 284–293. DOI:10.1016/S1474-4422(21)00001-6.

(9) Keeler, A. M.; Flotte, T. R. Recombinant Adeno-Associated Virus Gene Therapy in Light of Luxturna (and Zolgensma and Glybera): Where Are We, and How Did We Get Here? *Annual Review of Virology* **2019**, *6* (Volume 6, 2019), 601–621. DOI:10.1146/annurev-virology-092818-015530.

(10) Barenholz, Y. (Chezy). Doxil® — The First FDA-Approved Nano-Drug: Lessons Learned. *Journal of Controlled Release* **2012**, *160* (2), 117–134. DOI:10.1016/j.jconrel.2012.03.020.

(11) Sapsford, K. E.; Tyner, K. M.; Dair, B. J.; Deschamps, J. R.; Medintz, I. L. Analyzing Nanomaterial Bioconjugates: A Review of Current and Emerging Purification and Characterization Techniques. *Anal. Chem.* **2011**, *83* (12), 4453–4488. DOI:10.1021/ac200853a.

(12) Kondylis, P.; Schlicksup, C. J.; Zlotnick, A.; Jacobson, S. C. Analytical Techniques to Characterize the Structure, Properties, and Assembly of Virus Capsids. *Anal. Chem.* **2019**, *91* (1), 622–636. DOI:10.1021/acs.analchem.8b04824.

(13) Szatanek, R.; Baj-Krzyworzeka, M.; Zimoch, J.; Lekka, M.; Siedlar, M.; Baran, J. The Methods of Choice for Extracellular Vesicles (EVs) Characterization. *International Journal of Molecular Sciences* **2017**, *18* (6). DOI:10.3390/ijms18061153.

(14) Young, G.; Kukura, P. Interferometric Scattering Microscopy. *Annu. Rev. Phys. Chem.* **2019**, *70* (1), 301–322. DOI:10.1146/annurev-physchem-050317-021247.

(15) Desjardins, P.; Conklin, D. NanoDrop Microvolume Quantitation of Nucleic Acids. *J Vis Exp* **2010**, No. 45, 2565. DOI:10.3791/2565.

(16) Schmid, F.-X. Biological Macromolecules: UV-Visible Spectrophotometry. In *Encyclopedia of Life Sciences*; John Wiley & Sons, Ltd, 2001. DOI:10.1038/npg.els.0003142.

(17) Soltani, S.; Ojaghi, A.; Robles, F. E. Deep UV Dispersion and Absorption Spectroscopy of Biomolecules. *Biomed. Opt. Express, BOE* **2019**, *10* (2), 487–499. DOI:10.1364/BOE.10.000487.
(18) Bhat, S.; Curach, N.; Mostyn, T.; Bains, G. S.; Griffiths, K. R.; Emslie, K. R. Comparison of Methods for Accurate Quantification of DNA Mass Concentration with Traceability to the International System of Units. *Anal. Chem.* **2010**, *82* (17), 7185–7192. DOI:10.1021/ac100845m.
(19) Park, K. S.; Bae, Y. S.; Choi, S.-S.; Sohn, M. Y. High Numerical Aperture Reflective Deep Ultraviolet Fourier Ptychographic Microscopy for Nanofeature Imaging. *APL Photonics* **2022**, *7* (9), 096105. DOI:10.1063/5.0102413.
(20) Kumamoto, Y.; Taguchi, A.; Kawata, S. Deep-Ultraviolet Biomolecular Imaging and Analysis. *Advanced Optical Materials* **2019**, *7* (5), 1801099. DOI:10.1002/adom.201801099.
(21) Fang, X.; Polacchi, L.; Moreau, A.; Rigneault, H.; Bornschlögl, T.; Wenger, J. Direct Imaging of Single Gold and Polystyrene Nanoparticles in the Deep Ultraviolet. *Opt. Express, OE* **2025**, *33* (12), 24735–24745. DOI:10.1364/OE.558449.
(22) Zeskind, B. J.; Jordan, C. D.; Timp, W.; Trapani, L.; Waller, G.; Horodincu, V.; Ehrlich, D. J.; Matsudaira, P. Nucleic Acid and Protein Mass Mapping by Live-Cell Deep-Ultraviolet Microscopy. *Nat Methods* **2007**, *4* (7), 567–569. DOI:10.1038/nmeth1053.
(23) Cheung, M. C.; Evans, J. G.; McKenna, B.; Ehrlich, D. J. Deep Ultraviolet Mapping of Intracellular Protein and Nucleic Acid in Femtograms per Pixel. *Cytometry Part A* **2011**, *79A* (11), 920–932. DOI:10.1002/cyto.a.21111.
(24) Cheung, M. C.; LaCroix, R.; McKenna, B. K.; Liu, L.; Winkelman, J.; Ehrlich, D. J. Intracellular Protein and Nucleic Acid Measured in Eight Cell Types Using Deep-Ultraviolet Mass Mapping. *Cytometry Part A* **2013**, *83A* (6), 540–551. DOI:10.1002/cyto.a.22277.
(25) Cikaluk, B. D.; Masoumi, M. H.; Restall, B. S.; Martell, M. T.; Haven, N. J. M.; Zemp, R. J. Simultaneous Deep Ultraviolet Transmission and Scattering Microscopy for Virtual Histology. *Opt. Lett., OL* **2024**, *49* (10), 2729–2732. DOI:10.1364/OL.514077.
(26) Ojaghi, A.; Carrazana, G.; Caruso, C.; Abbas, A.; Myers, D. R.; Lam, W. A.; Robles, F. E. Label-Free Hematology Analysis Using Deep-Ultraviolet Microscopy. *Proceedings of the National Academy of Sciences* **2020**, *117* (26), 14779–14789. DOI:10.1073/pnas.2001404117.
(27) Wang, R.; Zhao, Q.; Quinn, J.; Yang, L.; Zhu, Y.; Huang, F.; Guo, C.; Wang, T.; Song, P.; Murphy, M.; Nguyen, T. D.; Maiden, A.; Robles, F. E.; Zheng, G. Deep-Ultraviolet Ptychographic Pocket-Scope (DART): Mesoscale Lensless Molecular Imaging with Label-Free Spectroscopic Contrast. *eLight* **2026**, *6* (1), 1. DOI:10.1186/s43593-025-00103-y.
(28) Gorti, V.; Kaza, N.; Williams, E. K.; Lam, W. A.; Robles, F. E. Compact and Low-Cost Deep-Ultraviolet Microscope System for Label-Free Molecular Imaging and Point-of-Care Hematological Analysis. *Biomed. Opt. Express, BOE* **2023**, *14* (3), 1245–1255. DOI:10.1364/BOE.482294.
(29) Soltani, S.; Ojaghi, A.; Qiao, H.; Kaza, N.; Li, X.; Dai, Q.; Osunkoya, A. O.; Robles, F. E. Prostate Cancer Histopathology Using Label-Free Multispectral Deep-UV Microscopy Quantifies Phenotypes of Tumor Aggressiveness and Enables Multiple Diagnostic Virtual Stains. *Sci Rep* **2022**, *12* (1), 9329. DOI:10.1038/s41598-022-13332-9.
(30) Yao, D.-K.; Maslov, K.; Shung, K. K.; Zhou, Q.; Wang, L. V. *In Vivo* Label-Free Photoacoustic Microscopy of Cell Nuclei by Excitation of DNA and RNA. *Opt. Lett., OL* **2010**, *35* (24), 4139–4141. DOI:10.1364/OL.35.004139.

(31) Wong, T. T. W.; Zhang, R.; Hai, P.; Zhang, C.; Pleitez, M. A.; Aft, R. L.; Novack, D. V.; Wang, L. V. Fast Label-Free Multilayered Histology-like Imaging of Human Breast Cancer by Photoacoustic Microscopy. *Science Advances* **2017**, *3* (5), e1602168. DOI:10.1126/sciadv.1602168.

(32) Fang, X.; Polacchi, L.; Zuttion, F.; Larquet, C.; Moreau, A.; Montaux-Lambert, A.; Rigneault, H.; Bornschlögl, T.; Wenger, J. Ultraviolet Extinction of Metal Oxide Nanoparticles Imaged at the Single Nanoaggregate Level. *Particle & Particle Systems Characterization* **2026**, *43* (4), e00180. DOI:10.1002/ppsc.202500180.

(33) Lombardo, C.; Sottini, A.; Seiter, S.; Colas des Francs, G.; Ortega Arroyo, J.; Quidant, R. Leveraging Partial Coherence to Enhance Nanoparticle Detection Sensitivity and Throughput in Interferometric Scattering Microscopy. *ACS Photonics* **2025**, *12* (8), 4376–4387. DOI:10.1021/acsphotonics.5c00744.

(34) Mazaheri, M.; Kasaian, K.; Albrecht, D.; Renger, J.; Utikal, T.; Holler, C.; Sandoghdar, V. iSCAT Microscopy and Particle Tracking with Tailored Spatial Coherence. *Optica, OPTICA* **2024**, *11* (7), 1030–1038. DOI:10.1364/OPTICA.523788.

(35) Zuo, C.; Sun, J.; Li, J.; Zhang, J.; Asundi, A.; Chen, Q. High-Resolution Transport-of-Intensity Quantitative Phase Microscopy with Annular Illumination. *Sci Rep* **2017**, *7* (1), 7654. DOI:10.1038/s41598-017-06837-1.

(36) Lemarchand, F. Private Communications (2013). Method Described in: DOI:10.1364/OE.20.015734.

(37) Gao, L.; Lemarchand, F.; Lequime, M. Refractive Index Determination of SiO2 Layer in the UV/Vis/NIR Range: Spectrophotometric Reverse Engineering on Single and Bi-Layer Designs. *J. Eur. Opt. Soc.-Rapid Publ.* **2013**, *8*, 13010. DOI:10.2971/jeos.2013.13010.

(38) Tian, L.; Waller, L. Quantitative Differential Phase Contrast Imaging in an LED Array Microscope. *Opt. Express, OE* **2015**, *23* (9), 11394–11403. DOI:10.1364/OE.23.011394.

(39) Effantin, G.; Boulanger, P.; Neumann, E.; Letellier, L.; Conway, J. F. Bacteriophage T5 Structure Reveals Similarities with HK97 and T4 Suggesting Evolutionary Relationships. *Journal of Molecular Biology* **2006**, *361* (5), 993–1002. DOI:10.1016/j.jmb.2006.06.081.

(40) Loofbourow, J. R.; Joyce, L. Increased Ultra–Violet Absorption of Cells Following Irradiation with Ultra–Violet Light. *Nature* **1941**, *148* (3745), 166–166. DOI:10.1038/148166a0.

(41) Kumamoto, Y.; Fujita, K.; Smith, N. I.; Kawata, S. Deep-UV Biological Imaging by Lanthanide Ion Molecular Protection. *Biomed. Opt. Express, BOE* **2016**, *7* (1), 158–170. DOI:10.1364/BOE.7.000158.

(42) Gorti, V.; McCubbins, K.; Houston, D.; Trenkle, A. D. S.; Holberton, A.; Serafini, C. E.; Wood, L.; Kwong, G.; Robles, F. E. Quantifying UV-Induced Photodamage for Longitudinal Live-Cell Imaging Applications of Deep-UV Microscopy. *Biomed. Opt. Express, BOE* **2025**, *16* (1), 208–221. DOI:10.1364/BOE.544778.

(43) Frutos, M. de; Letellier, L.; Raspaud, E. DNA Ejection from Bacteriophage T5: Analysis of the Kinetics and Energetics. *Biophysical Journal* **2005**, *88* (2), 1364–1370. DOI:10.1529/biophysj.104.048785.

(44) Leforestier, A.; Brasilès, S.; de Frutos, M.; Raspaud, E.; Letellier, L.; Tavares, P.; Livolant, F. Bacteriophage T5 DNA Ejection under Pressure. *Journal of Molecular Biology* **2008**, *384* (3), 730–739. DOI:10.1016/j.jmb.2008.09.035.

(45) Inagaki, T.; Hamm, R. N.; Arakawa, E. T.; Painter, L. R. Optical and Dielectric Properties of DNA in the Extreme Ultraviolet. *J. Chem. Phys.* **1974**, *61* (10), 4246–4250. DOI:10.1063/1.1681724.

(46) Leite, T. R.; Zschiedrich, L.; Kizilkaya, O.; McPeak, K. M. Resonant Plasmonic–Biomolecular Chiral Interactions in the Far-Ultraviolet: Enantiomeric Discrimination of Sub-10 Nm Amino Acid Films. *Nano Lett.* **2022**, *22* (18), 7343–7350. DOI:10.1021/acs.nanolett.2c01724.
(47) Piliarik, M.; Sandoghdar, V. Direct Optical Sensing of Single Unlabelled Proteins and Super-Resolution Imaging of Their Binding Sites. *Nat Commun* **2014**, *5* (1), 4495. DOI:10.1038/ncomms5495.
(48) Ortega Arroyo, J.; Andrecka, J.; Spillane, K. M.; Billington, N.; Takagi, Y.; Sellers, J. R.; Kukura, P. Label-Free, All-Optical Detection, Imaging, and Tracking of a Single Protein. *Nano Lett.* **2014**, *14* (4), 2065–2070. DOI:10.1021/nl500234t.
(49) Kashkanova, A. D.; Blessing, M.; Gemeinhardt, A.; Soulat, D.; Sandoghdar, V. Precision Size and Refractive Index Analysis of Weakly Scattering Nanoparticles in Polydispersions. *Nat Methods* **2022**, *19* (5), 586–593. DOI:10.1038/s41592-022-01460-z.
(50) Wu, J.; Zhang, Y.; Wu, Z.; Townsend, J.; Crooks, I.; Watt, B.; Baik, A.; Cygnar, K.; Qiu, H.; Li, N. Characterization of Lipid Nanoparticles Using Macro Mass Photometry: Insights into Size and Mass. *Analytica Chimica Acta* **2025**, *1351*, 343944. DOI:10.1016/j.aca.2025.343944.
(51) Wu, D.; Hwang, P.; Li, T.; Piszczek, G. Rapid Characterization of Adeno-Associated Virus (AAV) Gene Therapy Vectors by Mass Photometry. *Gene Ther* **2022**, *29* (12), 691–697. DOI:10.1038/s41434-021-00311-4.
(52) Wagner, C.; Fuchsberger, F. F.; Innthaler, B.; Lemmerer, M.; Birner-Gruenberger, R. Quantification of Empty, Partially Filled and Full Adeno-Associated Virus Vectors Using Mass Photometry. *International Journal of Molecular Sciences* **2023**, *24* (13). DOI:10.3390/ijms241311033.
(53) Liebel, M.; Hugall, J. T.; van Hulst, N. F. Ultrasensitive Label-Free Nanosensing and High-Speed Tracking of Single Proteins. *Nano Lett.* **2017**, *17* (2), 1277–1281. DOI:10.1021/acs.nanolett.6b05040.
(54) Brock, A. A.; Duraivel, S.; Jiang, J.; Fischer, J.; Greenberg, Z. F.; He, M.; Angelini, T. E.; Schmittgen, T. D. RNA Sequencing at Single Vesicle Resolution via 3D Printed Embedded Droplet Arrays. *ACS Appl. Mater. Interfaces* **2025**, *17* (38), 53110–53121. DOI:10.1021/acsami.5c09959.
(55) von Lersner, A. K.; Fernandes, F.; Ozawa, P. M. M.; Jackson, M.; Masureel, M.; Ho, H.; Lima, S. M.; Vagner, T.; Sung, B. H.; Wehbe, M.; Franze, K.; Pua, H.; Wilson, J. T.; Irish, J. M.; Weaver, A. M.; Di Vizio, D.; Zijlstra, A. Multiparametric Single-Vesicle Flow Cytometry Resolves Extracellular Vesicle Heterogeneity and Reveals Selective Regulation of Biogenesis and Cargo Distribution. *ACS Nano* **2024**, *18* (15), 10464–10484. DOI:10.1021/acsnano.3c11561.
(56) Prahl, S. *Mie Scattering Calculator*. DOI:10.5281/zenodo.10087653.
(57) Sumlin, B. J.; Heinson, W. R.; Chakrabarty, R. K. Retrieving the Aerosol Complex Refractive Index Using PyMieScatt: A Mie Computational Package with Visualization Capabilities. *Journal of Quantitative Spectroscopy and Radiative Transfer* **2018**, *205*, 127–134. DOI:10.1016/j.jqsrt.2017.10.012.
(58) Johnson, P. B.; Christy, R. W. Optical Constants of the Noble Metals. *Phys. Rev. B* **1972**, *6* (12), 4370–4379. DOI:10.1103/PhysRevB.6.4370.
(59) Birkhoff, R. D.; Painter, L. R.; Heller, J. M. Optical and Dielectric Functions of Liquid Glycerol from Gas Photoionization Measurements. *J. Chem. Phys.* **1978**, *69* (9), 4185–4188. DOI:10.1063/1.437098.

## Supplementary Information:

# Ultraviolet direct absorption microscopy for single particle protein/nucleic acid quantification

C. J. Richards,[1,*] D. Wolters,[1] D. van de Lockand,[1] and M. Liebel[1,$]

[1]Department of Physics and Astronomy, Vrije Universiteit Amsterdam, De Boelelaan 1100, Amsterdam 1081 HZ, The Netherlands.

[*] c.j.richards@vu.nl
[$] m.liebel@vu.nl

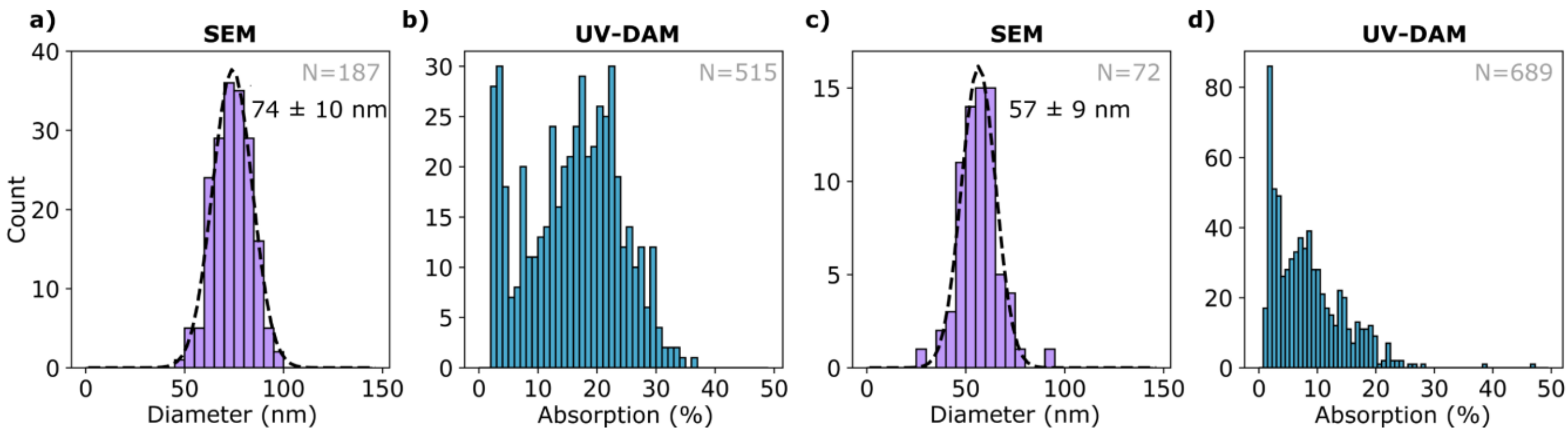

**Supplementary Figure S1. Au particle diameters and absorption values.** Au nanoparticles were imaged with SEM and UV-DAM (images shown in Figure 2). a) Distribution of diameters for nominally 60 nm Au particles based on SEM images. The particles have an average size of 74 ± 10 nm. Using Mie theory, this corresponds to an expected absorption value of 18%. More generally the diameters cover a range from 40–100 nm which corresponds to expected absorption values ranging 6–30%. b) Absorption value histogram based on UV-DAM imaging of the same sample as in a). Both the distribution center and range of absorption values are in good agreement with the expected values. c) Distribution of nominally 40 nm Au particle diameters based on SEM images. The particles have an average size of 57 ± 9 nm which corresponds to an expected absorption value of 11%. d) Absorption value histogram based on UV-DAM imaging of the same sample as in c). Likewise, the measured absorption values agree with the expected values. N denotes the number of particles used to make the distributions.

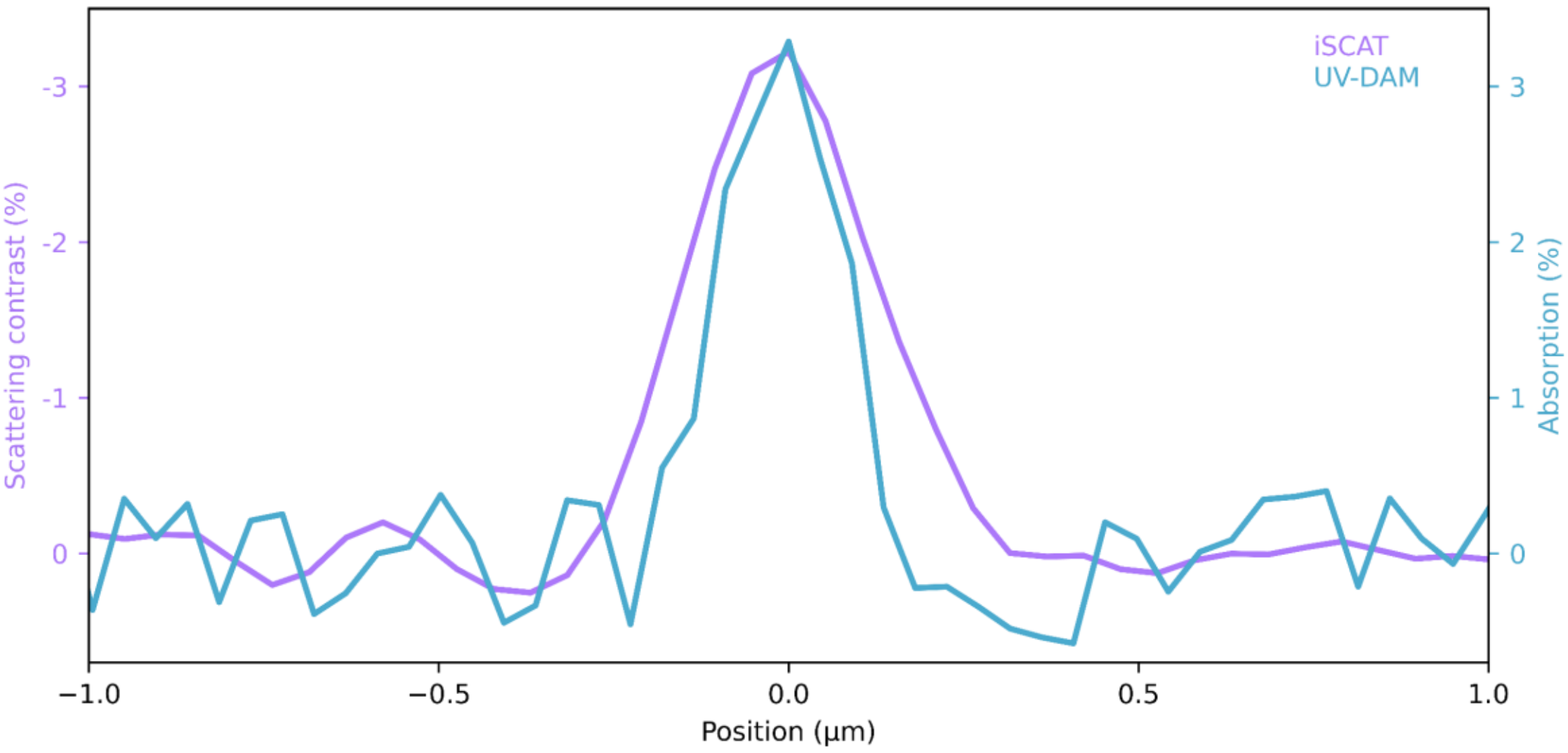

**Supplementary Figure S2. Phage line profiles.** Example line profiles of single T5 phages imaged with iSCAT using 470 nm light (purple) and UV-DAM using 280 nm illumination (blue). UV-DAM resolves the particles better than iSCAT due to the short illumination wavelength.

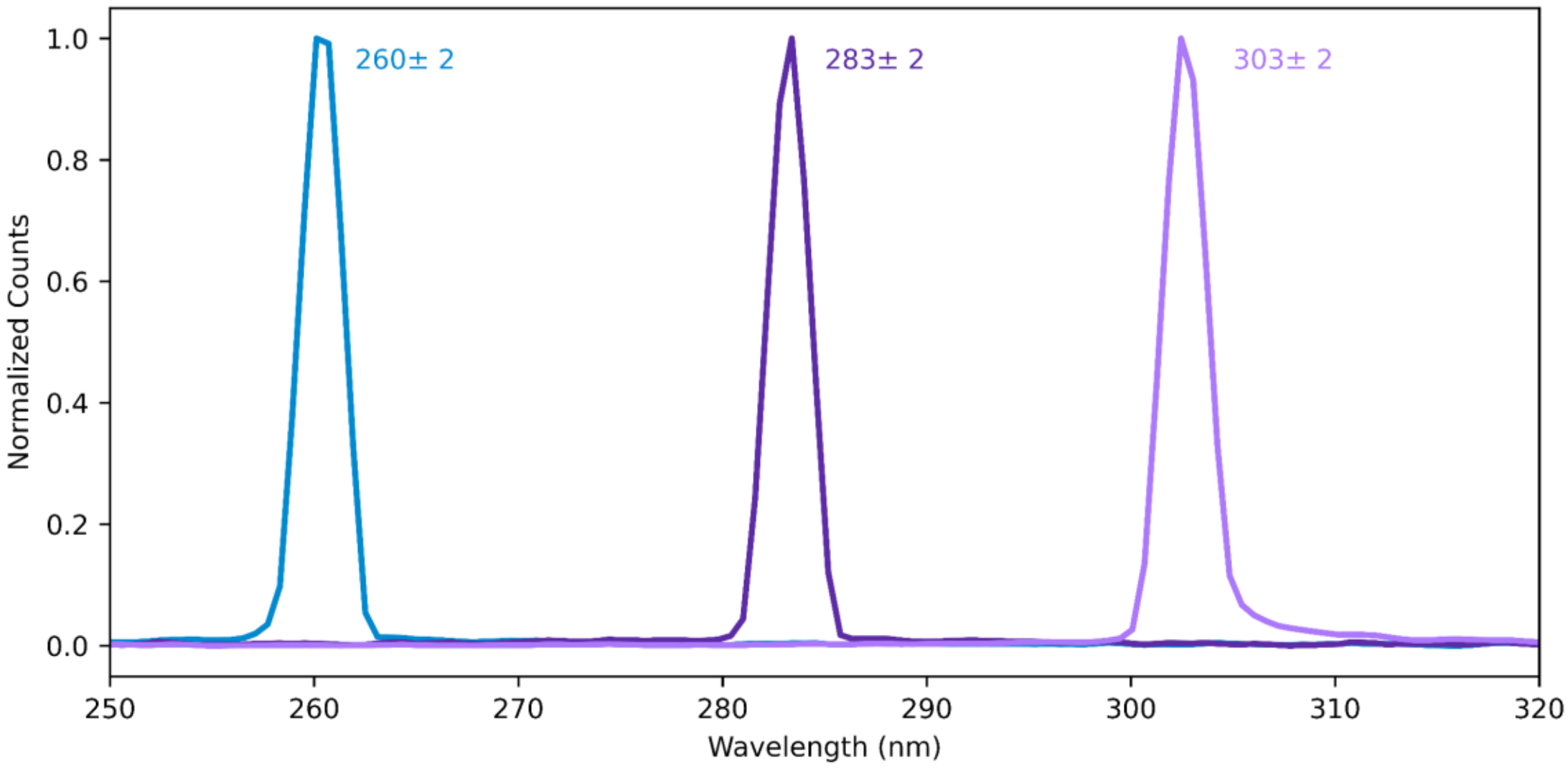

**Supplementary Figure S3. DUV source spectra.** Spectra of DUV light generated from the tunable source and used for imaging at ~260, 280, and 300 nm. Each spectra was normalized to a background reference and then fit with a gaussian. Means and standard deviations from the fits are listed.

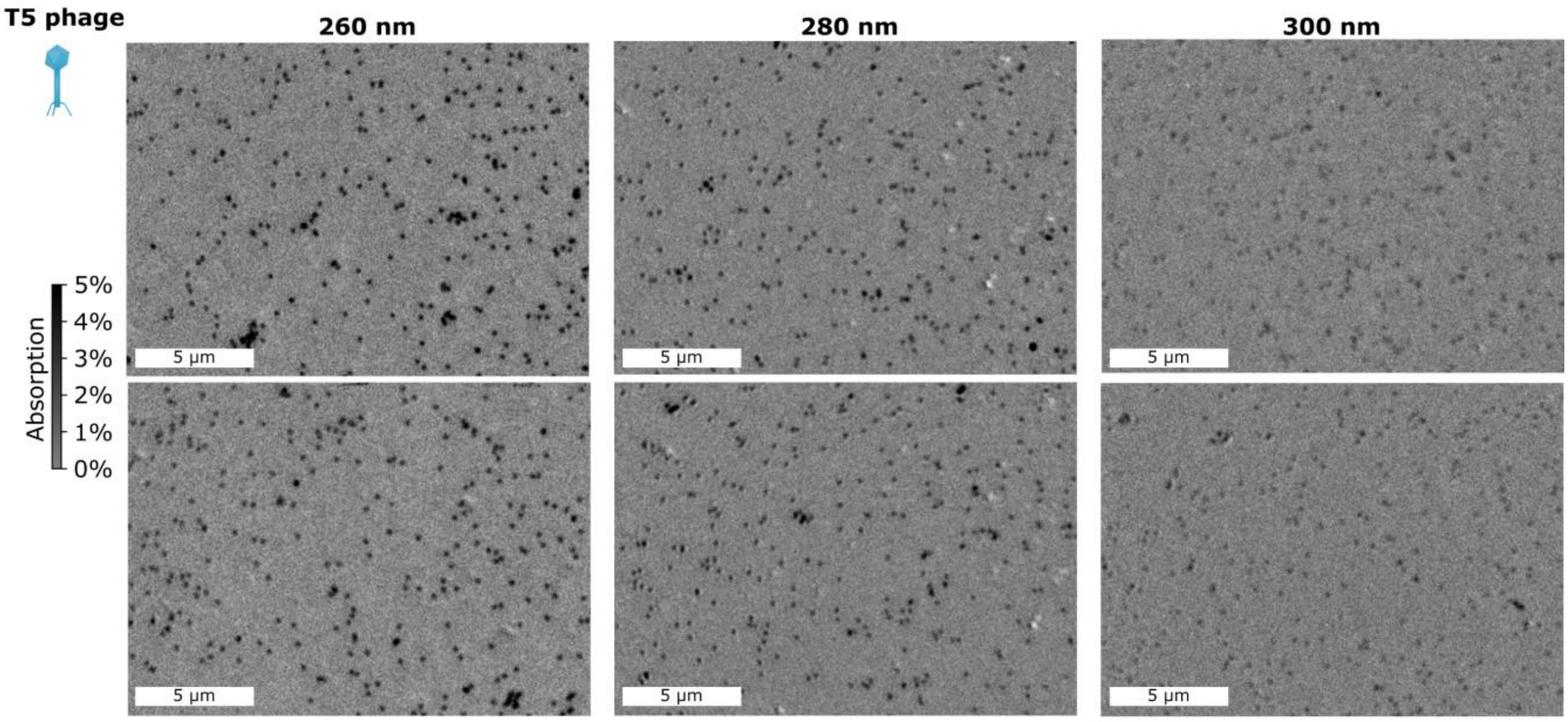

**Supplementary Figure S4. Additional DUV images of full T5 phages.** T5 phage samples imaged with UV-DAM using 260 nm, 280 nm, and 300 nm illumination. Individual particles are observed. Particle absorption values are similar within the sample, and decrease across the three illumination wavelengths.

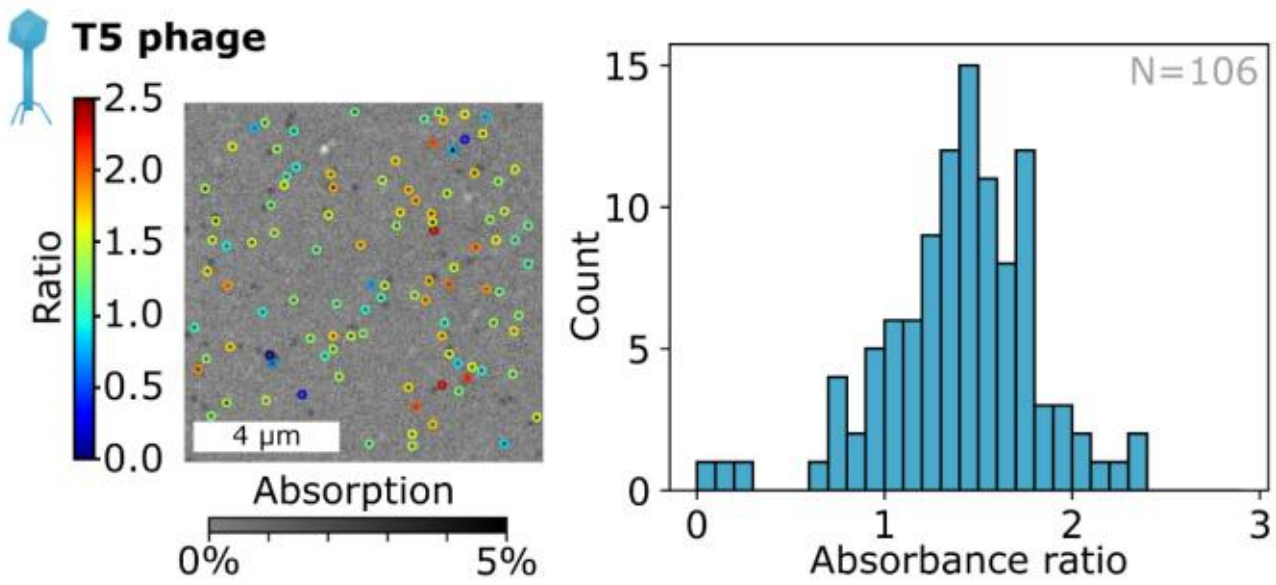

**Supplementary Figure S5. Single particle 260/280 nm absorbance ratios.** The same full T5 phages were imaged with both 260 nm and 280 nm illumination and individual particle absorption contrasts were obtained for both wavelengths. Left: UV-DAM image taken with 280 nm illumination. Circles indicate detected particles where the colour is the calculated individual particle 260/280 nm absorbance ratio. Right: corresponding histogram of the single particle absorbance ratios. The single particle absorbance ratios show a somewhat broad distribution with an average ratio of 1.4 ± 0.4, in good agreement with the UV-visible spectrometry value of 1.4. N denotes the number of particles used to form the distribution.

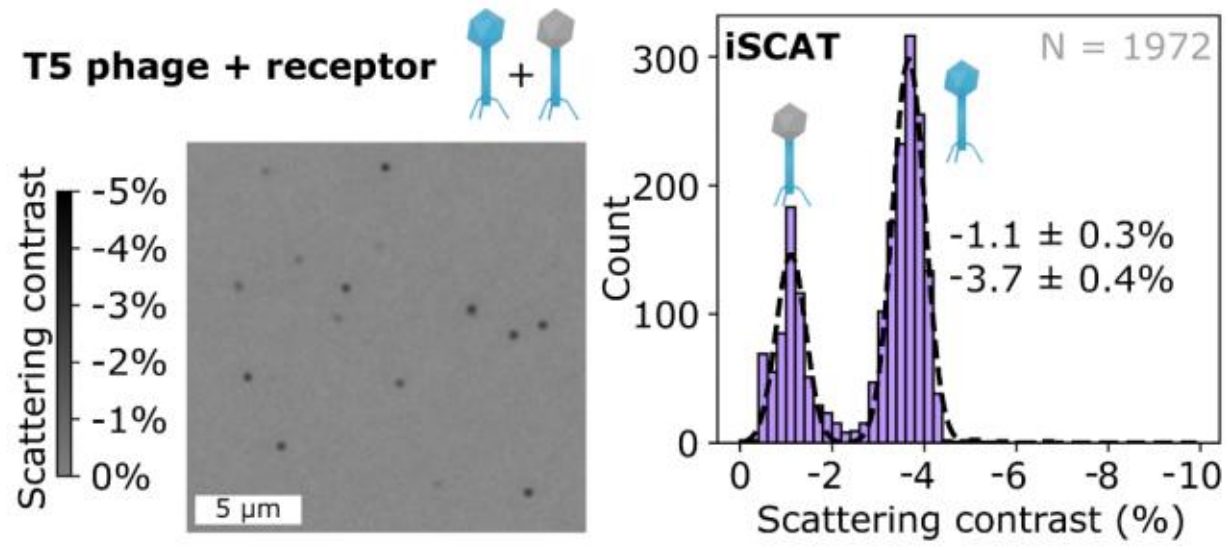

**Supplementary Figure S6. iSCAT measurements of mixed phage samples.** T5 phages were incubated with their receptor, FhuA, to produced mixed samples containing both full and empty capsids (schematically depicted as blue and grey capsids, respectively). Left: iSCAT images of mixed samples and right: corresponding scattering contrast histograms. Two clear populations are distinguishable pertaining to the lower scattering empty capsids and higher scattering contrast full capsids. The two component gaussian fit of the distribution is shown by the black dashed line and the corresponding means and standard deviations are listed. N denotes the number of particles used to form the distribution.